\documentclass[12pt]{article}
\usepackage{epsfig}
\usepackage{color}
\usepackage{amssymb,amsmath}
\usepackage{graphicx}

\newcommand{\bea}{\begin{eqnarray}}
\newcommand{\eea}{\end{eqnarray}}

 \makeatletter
\def\alt{\mathrel{\mathpalette\gl@align<}}
\def\agt{\mathrel{\mathpalette\gl@align>}}
\def\gl@align#1#2{\lower.6ex\vbox{\baselineskip\z@skip\lineskip\z@
\ialign{$\m@th#1\hfil##\hfil$\crcr#2\crcr\sim\crcr}}} \makeatother

\begin{document}
\begin{flushright}
UT-12-38
\end{flushright}
\vspace*{1.0cm}

\begin{center}
\baselineskip 20pt 
{\Large\bf 
Simple fermionic dark matter models \\
and Higgs boson couplings
}
\vspace{1cm}

{\large 
Nobuchika Okada$^{a}$ and Toshifumi Yamada$^{b}$
} \vspace{.5cm}

{\baselineskip 20pt \it
$^{a}$ Department of Physics and Astronomy, University of Alabama, \\
Tuscaloosa, Alabama 35487, USA \\

$^{b}$ Department of Physics, University of Tokyo,  \\
7-3-1 Bunkyoku, Tokyo 113-0033, Japan}

\vspace{.5cm}

\vspace{1.5cm} 
\large{\bf Abstract}
\end{center}

We consider a simple extension of the Standard Model (SM) 
 that incorporates a Majorana fermion dark matter 
 and a charged scalar particle with a  coupling 
 to the SM leptons through renormalizable terms.
Another renormalizable term involving the charged scalar 
 and the Higgs boson gives rise to interactions 
 between the dark matter and SM quarks at the one-loop level, 
 which induce the elastic scatterings between dark matter 
 and nucleus. 
The same term also affects the effective coupling 
 of the Higgs boson to diphoton through a one-loop diagram 
 with the charged scalar. 
Therefore, our model predicts a correlation between 
 the spin-independent cross section for dark matter-nucleus 
 elastic scatterings and a new contribution 
 to the effective Higgs boson coupling to diphoton. 
When the spin-dependent cross section is large enough
 to be tested in future direct dark matter detection experiments, 
 the Higgs-diphoton decay rate shows a sizable deviation 
 from the SM prediction. 
We also consider the case where the fermion dark matter is 
 a Dirac particle. 
Most of discussions is similar to the Majorana case, 
 but we find that the magnetic dipole moment of the Dirac fermion 
 dark matter is loop-induced and this interaction dominates 
 the spin-independent cross section 
 for dark matter-nucleus elastic scatterings. 
We find that the resultant cross section is about an order of 
 magnitude below the current experimental bound and 
 hence can be tested in the near future.  

\thispagestyle{empty}

\newpage

\addtocounter{page}{-1}
\setcounter{footnote}{0}
\baselineskip 18pt
%
\section{Introduction}

The Wilkinson Microwave Anisotropy Probe (WMAP) 
 observations \cite{wmap} have established 
 the existence of cold, non-baryonic dark matter (DM),
 whose abundance is measured to be
\begin{eqnarray}
\Omega_{\rm DM} h^2 = 0.112 \pm 0.006 \label{wmap} \ .
\end{eqnarray}
Since a DM candidate is missing in the standard model (SM),
 its existence motivates us to consider physics beyond the SM.
One notable fact about this measured DM abundance 
 is its possible relation to physics at the TeV scale.
Let us assume that the DM particle is a weakly-interacting 
 massive particle (WIMP)
 that has been in thermal equilibrium with SM particles and 
 decouples from the thermal bath when the temperature 
 drops below its mass scale,
 \textit{i.e.}, the DM particle is assumed to be a cold relic.
The measured DM abundance, Eq.~(\ref{wmap}), then implies that 
 the thermally averaged cross section times velocity
 for DM annihilations to SM particles 
 at the time of decoupling satisfies
\begin{eqnarray}
\langle \sigma v \rangle_{t_{dec}} \sim 
 \frac{ g^4 }{ {\rm TeV}^2 } \ ,
\end{eqnarray}
 with a coupling constant ($g$) between the DM and SM particles. 
Therefore, if the coupling constant is of the order 
 of the electroweak gauge couplings, 
 dark matter physics is controlled by the TeV scale.

Motivated by the fact above,
 we study a simple extension of the SM at the TeV scale that
 incorporates a DM particle and its interaction with SM particles.
In this paper, we try to extend the SM as minimally as possible.
Models for DM have often been associated with solutions 
 to the gauge hierarchy problem, a well-known example 
 of which is the neutralino DM scenario 
 in minimal supersymmetric standard model.
However, the ATLAS and the CMS experiments 
 searching for supersymmetric particles 
 at the Large Hadron Collider (LHC) 
 have been reporting negative results in all channels,
 that might cast doubts on the notion of naturalness 
 and the validity of the gauge hierarchy problem. 
In this paper, we adopt the ``minimality" as the guiding principle 
 for model-building, instead of the naturalness of the electroweak scale.

The simplest renormalizable extension of SM that incorporates 
 a DM particle is the so-called Higgs portal model \cite{HP},
 where a SM gauge singlet scalar is introduced as a DM particle 
 and it interacts only with the Higgs boson
 through a renormalizable term.
Unfortunately, this model is in tension with XENON 100
 direct DM detection experiment \cite{xenon100},
 except for limited parameter regions where the DM mass is 
 very close to half of the Higgs boson mass,
 or the DM mass is larger than about 200 GeV \cite{portal}.
%
Although the Higgs portal dark matter might be discovered 
 by future direct DM search experiments, 
 the current allowed region is virtually impossible
 to test at collider experiments \cite{portal};
when the DM mass is larger than half of the Higgs boson mass, 
 the DM particles are produced only through off-shell Higgs boson and 
 the production cross section is extremely small.
The DM particles are produced through the Higgs boson decay 
 when the DM mass is smaller than half of the Higgs boson mass.
In the allowed region, however, the coupling constant among Higgs bosons and DM particles
 are so small
 that the branching ratio of the invisible Higgs decay is too tiny to observe.

Here we consider the second simplest renormalizable extension 
 of the SM with a fermionic DM particle.
In this model, we introduce a SM gauge singlet fermion ($\chi$)  
 and a scalar ($S$) charged under SM gauge groups. 
Odd $Z_2$ parity is assigned to $S$ and $\chi$, 
 whereas even parity is assigned to all the SM particles.
Hence, assuming $\chi$ is lighter than $S$, 
 the lightest $Z_2$ odd particle, $\chi$, is stable 
 and a DM candidate.
$S$ and $\chi$ can interact with SM particles 
 through the following renormalizable term:
\begin{eqnarray}
{\cal L} \supset - y_i \ \bar{\psi}_i S \chi \ + \ {\rm h.c.} \ , 
\label{coupling1}
\end{eqnarray}
 where $\psi_i$ denotes a SM particle with flavor index $i$
 that has the same SM charge as $S$.
If the coupling constant $y_i$ is of the order 
 of the electroweak gauge couplings, 
 then the measured DM abundance in Eq.~(\ref{wmap}) 
 implies that the particles, $\chi$ and $S$,  
 have the mass of order 100 GeV$- 1$ TeV.
There is a variety of choices for the charge assignment 
 of $S$ to allow the interaction in Eq.~(\ref{coupling1}). 
In this paper, we consider the case 
 that $S$ is color-singlet and has the mass below 400 GeV,
 and the mass difference between $S$ and $\chi$ are smaller 
 than 50 GeV.
Such a parameter choice is interesting 
 because the charged scalar $S$ has rich phenomenological implications, 
 especially on the branching ratio of the Higgs boson into diphoton,
 but it will take time for the LHC experiments 
 to exclude or discover the existence of the scalar and DM particles.

The heart of this study lies in the fact that 
 since $S$ is a scalar particle, no symmetry forbids 
 the following renormalizable term:
\begin{eqnarray}
{\cal L} \supset - \kappa (S^{\dagger} S) (H^{\dagger} H) \ ,
\end{eqnarray}
 where $H$ is the Higgs boson in the SM.
This term induces at the one-loop level an interaction
 between $\chi$ and SM quarks mediated by the Higgs boson 
 and the spin-independent cross section of $\chi$-nucleus 
 elastic scatterings can be accessible by future 
 direct DM detection experiments.
On the other hand, since $S$ has an electric charge 
 and a mass less than several 100 GeV,
 the same term contributes to the Higgs boson decay 
 to diphoton through a loop diagram, 
 altering the Higgs-to-diphoton branching ratio
 from its SM prediction. 
Our model thus predicts both a signal in direct DM detection 
 experiments and a deviation of the Higgs-to-diphoton 
 branching ratio.

Another important result of this study is that 
 if the DM particle is a Dirac fermion, 
 the magnetic dipole moment of the DM particle 
 is induced at the one-loop level. 
This interaction gives a dominant contribution 
 to the DM-nucleus elastic scatterings. 
We will find that the spin-independent cross section 
 for the processes is below the current experimental bounds 
 but within the reach of future experiments.

This paper is organized as follows.
In the next section, we define our model. 
In Sec.~3, we investigate DM physics with our model. 
By calculating the relic abundance of the DM particle 
 and comparing our results to the observations, 
 we identify allowed parameter regions of the model.
We also calculate the spin-independent cross section 
 for the elastic scatterings between the DM particle and nucleus
 mediated by the Higgs boson. 
In Sec.~4, the effect of the charged scalar, $S$, 
 on the Higgs branching ratio to diphoton is calculated.
Then, in Sec.~5, we show correlations between the DM-nucleus 
 elastic scattering cross section and the Higgs-to-diphoton 
 branching ratio. 
Finally in Sec.~6, we discuss the case for 
 a Dirac fermion DM particle and investigate 
 its phenomenology.

\section{Simple model for Majorana fermion dark matter}
We consider a simple renormalizable extension of the SM 
 which incorporates a fermionic DM particle
 interacting with SM particles.
We introduce a SM gauge singlet Majorana fermion ($\chi$)
 and a complex scalar field ($S$) with a hypercharge $Y=-1$. 
Odd-parity is assigned to $\chi$ and $S$, 
 while even-parity to all the SM fields.
The renormalizable terms involving $\chi$ and $S$ are the following:
\begin{eqnarray}
{\cal L}_{extra} &=& \frac{i}{2} \ \bar{\chi} \gamma^\mu \partial_\mu \chi 
  - \frac{1}{2} \ m_{\chi} \bar{\chi} \chi 
  + (D^{\mu} S)^{\dagger} (D_{\mu} S) 
  - m_{0}^2 S^{\dagger}S  - \lambda_S \ (S^{\dagger}S)^2
\nonumber \\
&-& y_i \ \bar{e}_{iR}  S  \chi  +  {\rm h.c.} 
   - \kappa (S^{\dagger}S)(H^{\dagger} H)  \ ,
\label{basicL}
\end{eqnarray}
 where $e_{Ri}$ denotes the SM SU(2)$_L$ singlet charged lepton 
 with flavor index $i$, $H$ denotes the SM Higgs doublet field, 
 and $y_i$ and $\kappa$ are dimensionless couplings 
 between the new particles and the SM fields.
The field contents are summarized in Table 1.
\begin{table}
\begin{center}
\begin{tabular}{|c|c|c|c|} \hline
Field
& SU(2)$_L$
& U(1)$_Y$
& $Z_2$
\\ \hline
$e_{Ri}$    & ${\bf 1}$ & $-1$   & $+$ \\ \hline
$H$         & ${\bf 2}$ & $-1/2$ & $+$ \\ \hline
$\chi$      & ${\bf 1}$ & $0$    & $-$   \\ \hline
$S$         & ${\bf 1}$ & $-1$   & $-$  \\ \hline
\end{tabular}
\end{center}
\caption{The field content of the model}
\end{table}
The DM particle $\chi$ is assumed to be lighter than $S$, 
 so that $\chi$ is stable due to the $Z_2$ parity 
 and can be a DM candidate.

For general values of $y_i$'s, the term $y_i \bar{e}_{iR} S \chi$ 
 induces flavor-changing neutral current processes,
 which are severely constrained by current experimental results. 
Here we simply assume that $y_i$'s are aligned 
 with respect to the SM lepton flavors,
 i.e., we set $y_i = y \delta_{ei}, \ y \delta_{\mu i}$
 or $y \delta_{\tau i}$.

The masses of $\chi$ and $S$ are, respectively, given by
 $m_{\chi}$ and $m_{0}^2 + \kappa v^2/2 \equiv m_S^2$, 
 after the electroweak symmetry breaking 
 by $\langle H \rangle = (v,0)^T$ with $v=174$ GeV. 
Our model is possibly constrained by the direct slepton search 
 performed by the ATLAS experiment \cite{lhc slepton}.
We here take $m_S > 100$ GeV and $m_S - m_{\chi} < 50$ GeV, 
 so that the model safely evades the ATLAS bound 
 as well as the LEP bound on charged sleptons \cite{lep}.

\section{Dark matter physics}
In this section, we constrain the model parameters
 from the requirement that $\chi$ be a cold relic
 whose abundance fits with the WMAP data of Eq.~(\ref{wmap}).
We then calculate the cross section for the DM-nucleus 
 elastic scatterings and compare it with 
 the current experimental bounds.

\subsection{Relic abundance}
The DM particle $\chi$ interacts with a SM charged 
 lepton through the coupling $y_i$.
When the temperature of the radiation-dominated Universe 
 cools down below $T \sim m_{\chi}$,
 $\chi$ decouples from the thermal bath and its abundance freezes out.
The decoupling process of $\chi$ is regulated 
 by the Boltzmann equation \cite{kolb turner}.
Let $s$, $n_{\chi}$ and $n^{{\rm eq}}_{\chi}$, respectively,
 denote the entropy density, the number density of $\chi$ 
 and the number density if $\chi$ were to stay in thermal equilibrium.
We further define the following quantities:
\begin{eqnarray}
z &\equiv& \frac{m_{\chi}}{T} \ ,
\\
Y(z) &\equiv& \frac{ n_{\chi}(z) }{ s } \ ,
\\
Y^{{\rm eq}}(z) &\equiv& \frac{ n^{{\rm eq}}_{\chi}(z) }{ s } \ ,
\\
H(T) &\equiv& \sqrt{\frac{\pi^2}{90} g_*} \frac{T^2}{M_P} \ ,
\end{eqnarray}
 where $g_*$ denotes the effective thermal degrees of freedom
 during the decoupling process of $\chi$,
 and $M_P=2.4 \times 10^{18}$ GeV is the reduced Planck mass.
For $m_{\chi} \lesssim 300$ GeV, we have $g_* = 86.25$.
For $z \gtrsim 3$, $n^{{\rm eq}}_{\chi}(z)$ can be written as
\begin{eqnarray}
n^{{\rm eq}}_{\chi}(z) &=& g_{\chi} \frac{T m_{\chi}^2}{2 \pi^2} K_2(z) \ ,
\end{eqnarray}
 where $g_{\chi}=2$ counts the physical degree of freedom 
 of the Majorana fermion $\chi$.

We consider the DM pair annihilation process  
$ \chi \ \chi \rightarrow e_i \ \bar{e}_i$, 
 for which the thermally averaged cross section 
 times velocity is expressed as \cite{relic density} 
\begin{eqnarray}
\langle \sigma v \rangle 
&=& g_{\chi}^2 \ \frac{1}{ ( n^{{\rm eq}}_{\chi}(z))^2 } \ \frac{T}{32 \pi^4} \ \int^{\infty}_{4 m_{\chi}^2} {\rm d}s \ 
\sqrt{ \frac{s}{4} - m_{\chi}^2 } \ K_1 \left( \frac{\sqrt{s}}{T} \right) \ \vert {\cal M}(s) \vert^2 \ ,
\end{eqnarray}
 where $\vert {\cal M}(s) \vert^2$ denotes the scattering rate 
 per co-moving volume for the process and is given by
\begin{eqnarray}
\vert {\cal M}(s) \vert^2 &=& \frac{1}{g_{\chi}^2} \ \frac{y^4}{16 \pi} \ 4 \
\left\{ (m_S^2 - m_{\chi}^2)^2 + \frac{s}{2} \ m_S^2 \right\} \nonumber \\
&\times& \frac{ \ \sqrt{s} \sqrt{ \frac{s}{4} - m_{\chi}^2 } \left( \frac{s}{2} - m_{\chi}^2 + m_S^2 \right)
\ - \ \left\{ (m_S^2 - m_{\chi}^2)^2 + s \, m_S^2 \right\} \arctan \left( \frac{ \sqrt{s} \sqrt{s/4 - m_{\chi}^2} }{ s/2 - m_{\chi}^2 + m_S^2 } \right)
\ }
{ \ \frac{ \sqrt{s} }{2} \sqrt{ \frac{s}{4} - m_{\chi}^2 } \ \left( \frac{s}{2} - m_{\chi}^2 + m_S^2 \right) \ 
\left\{ (m_S^2 - m_{\chi}^2)^2 + s \, m_S^2 \right\} \ } \ . \nonumber \\
\end{eqnarray}
The Boltzmann equation takes the form:
\begin{eqnarray}
\frac{{\rm d} Y(z)}{{\rm d} z} &=& 
- \frac{z \ \langle \sigma v \rangle \ s}{H(m_{\chi})} 
 \ ( \ Y^2(z) - Y_{{\rm eq}}^2(z) \ ) \ .
\label{boltz}
\end{eqnarray}

We numerically solve the Boltzmann equation 
 for various values of $y$ and $m_S$.
In our analysis, the mass of $\chi$ is taken 
 so as to satisfy the relations, $m_{\chi} = m_S-50$ GeV 
 and $m_{\chi} = m_S-20$ GeV, for simplicity. 
We evaluate the relic abundance of $\chi$ as $Y(z=\infty)$. 
In Fig.~1, 
 we show contour plots of the relic abundance of $\chi$
 on the ($m_S$, $y^2$)-plane. 
Regions between two (red) lines correspond to the parameter regions 
 where the resultant DM relic abundance fits with
 the observed values in Eq. (\ref{wmap}).
 From Fig.~1, we can see that the mass difference 
 between $\chi$ and $S$ does not significantly 
 alter the relic abundance as long as 
 the difference is within $O(10 \; {\rm GeV})$. 
For the left panel in Fig.~1, 
 we have found that the following relation must be satisfied 
 in order to reproduce the observed DM abundance:
\begin{eqnarray}
y^2  \simeq a  ( m_S / {\rm GeV}  -  100  )  +  b \ , 
\label{y-ms}
\end{eqnarray}
 with 
\begin{eqnarray}
0.0038 \leq a \leq 0.0040, \ \ \ 0.34 \leq b \leq 0.37 \ .
\end{eqnarray}

\begin{figure}[htbp]
 \begin{minipage}{0.5\hsize}
  \begin{center}
   \includegraphics[width=80mm]{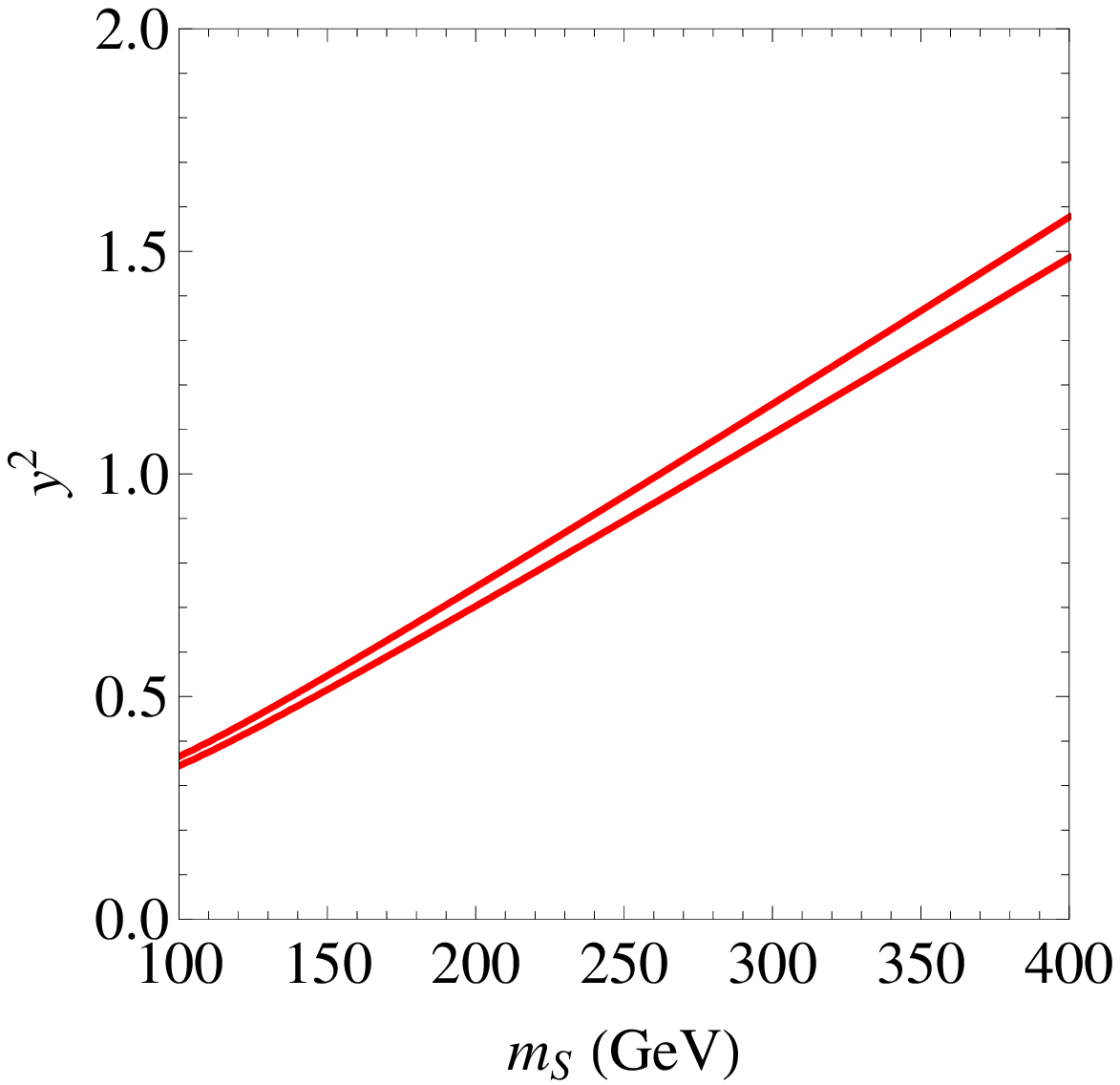}
  \end{center}
 \end{minipage}
 \begin{minipage}{0.5\hsize}
  \begin{center}
   \includegraphics[width=80mm]{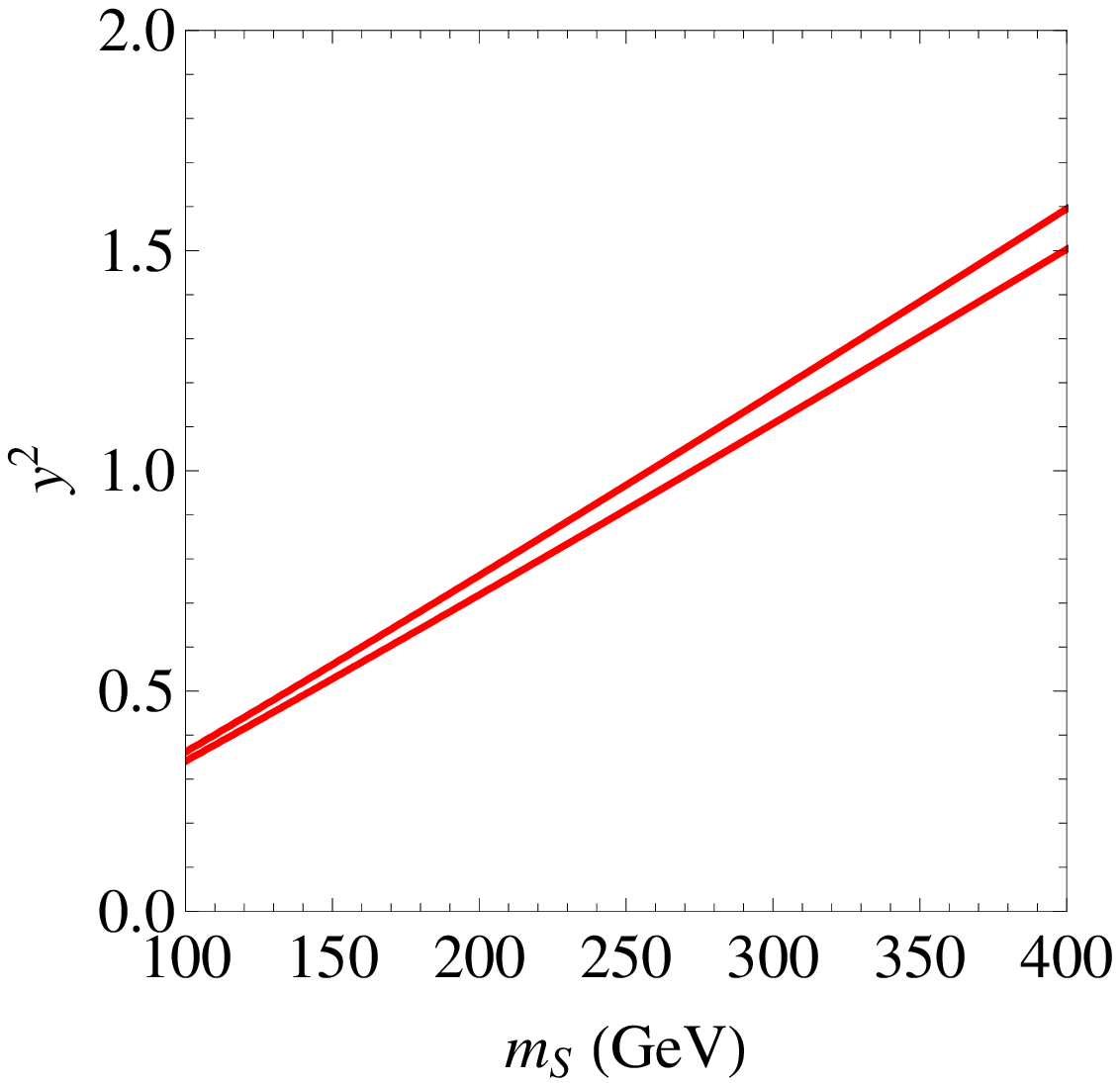}
  \end{center}
 \end{minipage}
\caption{
Contour plot for the relic abundance of the DM candidate $\chi$
 on the ($m_S$, $y^2$)-plane. 
We have taken $m_{\chi}=m_S - 50$ GeV ($m_{\chi}=m_S - 20$ GeV) 
 in the left (right) panel. 
The region between two (red) lines corresponds to 
 the range of the observed DM abundance in Eq.~(\ref{wmap}).
 }
\end{figure}

\subsection{Direct Dark Matter Detection}
The DM particle $\chi$ couples to SM quarks 
 at the one-loop level through the Feynman diagrams 
 in Fig.~3\footnote{
Since $\chi$ is a Majorana fermion, 
 photon exchange does not contribute,
 as it only gives rise to a magnetic dipole moment coupling.
}.
We calculate effective couplings between $\chi$ and a SM quark.
The Higgs boson exchange leads to 
\begin{eqnarray}
{\cal L}_{\rm eff}^{h} = m_Q (\bar{Q}Q) (\bar{\chi}\chi) 
 \left( \frac{-1}{16 \pi^2} \right)
  \kappa \ y^2 \ \frac{1}{m_h^2} 
\ \frac{m_{\chi}}{m_S^2} \ \frac{ x + (1-x) \ln(1-x) }{ x^2 } \ , \label{hex}
\end{eqnarray}
 where $Q$ denotes a SM quark and $x \equiv m_{\chi}^2/m_S^2$.
An effective coupling that arises from the $Z$-boson exchange
 is given by
\begin{eqnarray}
{\cal L}_{\rm eff}^{Z} &=& m_Q (\bar{Q} \gamma_5 Q) 
(\bar{\chi} \gamma_5 \chi) \nonumber \\
&\times& \frac{1}{16 \pi^2} \
g_{Z \ell} \ g_{ZQ_A} \ y^2 \ \frac{1}{M_Z^2} 
\left( \frac{m_{\chi}}{m_S^2} \right)
\frac{ \left( 1-\frac{x}{3} \right) \ \ln (1-x) \ 
 - \ x \left\{ 1 + \frac{2}{3} \ln \left( \frac{m_\ell^2}{m_S^2}
\right)  \right\}}{ 2(1-x)x } \ , 
\label{zex}
\end{eqnarray}
 where $m_\ell$ denotes the mass of the SM lepton, 
 $g_{Z \ell}$ represents the coupling 
 between the $Z$-boson and the SM lepton, 
 and $g_{ZQ_A}$ denotes the axial part of the coupling 
 between $Z$-boson and the quark $Q$.
In deriving Eqs.~(\ref{hex}) and (\ref{zex}), 
 we have taken the limit $m_\ell \ll m_\chi$.

\begin{figure}[htbp]
  \begin{center}
   \includegraphics[width=130mm]{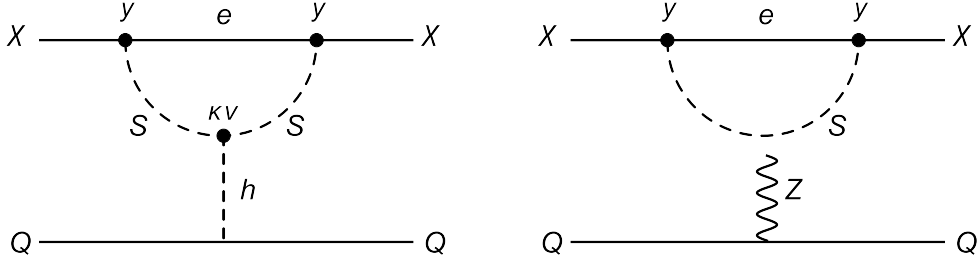}
  \end{center}
 \caption{
Feynman diagrams contributing to the interaction 
 between $\chi$ and a SM quark. 
The left diagram represents the Higgs boson exchange 
 and the right one the $Z$-boson exchange.
 }
\end{figure}

Since the coupling between $\chi$ and a SM quark induced 
 by the $Z$-boson exchange is of pseudo-scalar type,
 its contribution to DM-nucleus elastic scatterings
 is spin-dependent and is proportional to
 a DM velocity squared in the Earth frame. 
Therefore this contribution is negligibly small. 
On the other hand, the coupling induced by the Higgs boson exchange 
 diagram is of scalar-type and hence spin-independent.
This contribution to the cross section of DM-nucleus 
 elastic scatterings can be accessible 
 by direct DM detection experiments 
 if a value  of $\kappa$ is large enough. 
In the following, we only consider the contribution 
 from the Higgs boson exchange diagram.

\begin{figure}[t]
  \begin{center}
   \includegraphics[width=120mm]{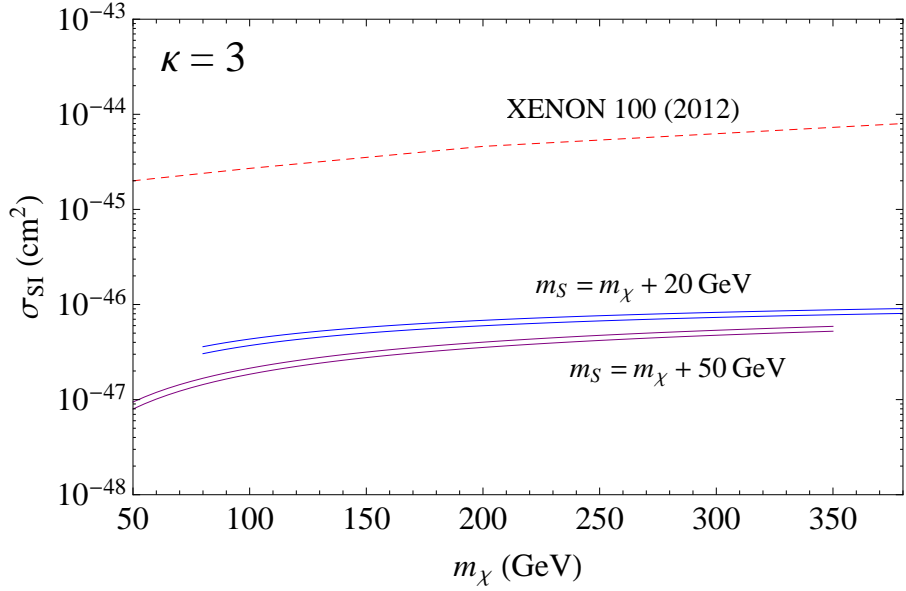}
  \end{center}
 \caption{
The spin-independent cross section for DM-nucleon elastic scattering
 as a function of the DM mass, $m_{\chi}$.
Here we have fixed $\kappa=3$. 
The charged scalar mass, $m_S$, is related to the DM mass as
 $m_{\chi}=m_S-50$ GeV (lower two solid lines)
 and $m_{\chi}=m_S-20$ GeV (upper two solid lines), respectively. 
The coupling constant $y$ is taken to satisfy the relation
  of Eq.~(\ref{y-ms2}) so that the model reproduces 
 the observed DM abundance.
For each pair of degenerate lines, the upper one corresponds 
 to $(a=0.0040, \ b=0.37)$, while the lower one to 
 $(a=0.0038, \ b=0.34)$ in eq.~(\ref{y-ms}).
The current experimental bound reported 
 by the XENON 100 experiment \cite{xenon100} 
 is shown by the dashed line.
 }
\end{figure}

Let us calculate the cross section for DM-nucleon elastic scattering.
We first estimate the following matrix elements:
\begin{eqnarray}
\langle N \vert \ m_Q \bar{Q}Q \ \vert N \rangle
 \equiv m_N f_{TQ}^N \ \ \ (Q=u,d,s,c,b,t) \ ,
\end{eqnarray}
 where $N$ represents a proton, $p$, or a neutron, $n$.
A recent lattice calculation \cite{lattice} reports that
\begin{eqnarray}
f_{Tu}^p + f_{Td}^p = f_{Tu}^n + f_{Td}^n \simeq 0.056 , 
  \ \ \ \vert f_{Ts}^N \vert \leq 0.08 \ .
\label{ftud}
\end{eqnarray}
We hereafter adopt these values, but we set $f_{Ts}^N=0$ 
 to make our analysis conservative. 
Operators involving heavy quarks, $c,b,t$, 
 give rise to an effective coupling to gluons
 through a triangle diagram \cite{heavy quark contributions},
 and in heavy quark limit we have
\begin{eqnarray}
\langle N \vert \ m_Q \bar{Q}Q \ \vert N  \rangle
 \simeq  
\langle N \vert \ \left( - \frac{ 1 }{ 12 \pi } \alpha_s \right) G_{\mu \nu} G^{\mu \nu} \ \vert N \rangle \ \ \ {\rm for} \ Q=c,b,t.
\end{eqnarray}
Using the trace of the QCD energy momentum tensor given by 
\begin{eqnarray}
\theta^{\mu}_{\mu} = \sum_{Q=u,d,s,c,b,t} m_Q \bar{Q}Q 
  - \frac{ 7 \alpha_s }{8 \pi} G_{\mu \nu}G^{\mu \nu}
 \simeq  \sum_{Q=u,d,s} m_Q \bar{Q}Q  
 -  \frac{ 9 \alpha_s }{8 \pi} G_{\mu \nu}G^{\mu \nu} \ ,
\end{eqnarray}
 we obtain 
\begin{eqnarray}
m_N \simeq 
\langle N \vert \sum_{Q=u,d,s} m_Q \bar{Q}Q  \vert N \rangle
  +   \langle N \vert  \left( \frac{ -9 \alpha_s }{ 8 \pi } \right) 
 G_{\mu \nu} G^{\mu \nu}  \vert N \rangle \ .
\end{eqnarray}
We thus find for $Q=c,b,t$,
\begin{eqnarray}
\langle N \vert \ m_Q \bar{Q}Q \ \vert N \rangle &\simeq& 
\langle N \vert \ \left( \frac{ -2 \alpha_s }{ 24 \pi } \right) G_{\mu \nu} G^{\mu \nu} \ \vert N \rangle \nonumber \\
&\simeq& \frac{2}{27} \ ( \ 1 - f_{Tu}^N - f_{Td}^N - f_{Ts}^N \ ) \ m_N \ .
\end{eqnarray}
Using above expressions, the spin-independent cross section 
 for the elastic scattering of $\chi$ with a nucleon is given by 
\begin{eqnarray}
\sigma_{\rm SI} &=& \frac{1}{\pi} \ \frac{m_{N}^2 \, m_{\chi}^2}{(m_{N}+m_{\chi})^2} \ 
m_{N}^2 \ 
\left\{ \ f_{Tu}^N + f_{Td}^N + f_{Ts}^N +
\frac{2}{9} ( 1 - f_{Tu}^N - f_{Td}^N - f_{Ts}^N ) \ \right\}^2 \ \nonumber
\\
&\times& \left\{ \ \frac{1}{16 \pi^2} \ \kappa \ y^2 \ \frac{1}{m_h^2} 
\ \frac{m_{\chi}}{m_S^2} \ \frac{ x + (1-x) \ln(1-x) }{ x^2 } \ \right\}^2 \ ,
\label{sigmaSI}
\end{eqnarray}
 where $m_{N} \simeq 1$ GeV is the mass of a nucleon (proton or neutron), 
 and $x = m_{\chi}^2/m_S^2$.

We evaluate $\sigma_{\rm SI}$ for various values of 
 $m_{\chi}=m_S-50$ GeV and $m_{\chi}=m_S-20$ GeV, respectively. 
Here, $y$ is given as a function of $m_S$ 
 by using Eq.~(\ref{y-ms}) so as to reproduce the observed 
 dark matter relic abundance, 
 and we fix the Higgs boson mass as $m_h=125$ GeV. 
In Fig.~3, we show $\sigma_{\rm SI}$ as a function of $m_{\chi}$ 
 for $\kappa=3$, along with the current upper bound 
 reported by the XENON 100 experiment \cite{xenon100}.  
The resultant cross section is found to be 
 (far) below the upper bound. 
Note that $\sigma_{\rm SI}$ scales as $\kappa^2$. 
If we allow a value of $|\kappa|$ 
 as large as its naive upper bound, $|\kappa| \leq 4 \pi$, 
 from the view point of perturbation theory, 
 the spin-independent cross section can be close to 
 the current limit. 
The sensitivity of the spin-independent cross section 
 in the future direct DM detection experiments, 
 such as the XENON 1T \cite{xenon 1t}, 
 is expected to reach $(2-8) \times 10^{-47}$ cm$^{-2}$, 
 and hence our scenario can be tested.
As we will discuss in the next sections, 
 the spin-independent cross section has a correlation 
 with a deviation of the Higgs-to-diphoton branching 
 ratio from the SM predicted value. 
We will see that the deviation becomes larger 
 as $|\kappa|$ is raised, equivalently, 
 the spin-independent cross section becomes larger.

\section{Anomalous Higgs coupling to diphoton}

Although the Higgs boson discovered 
 at the LHC experiments \cite{lhc higgs} 
 has shown its properties mostly consistent 
 with those expected in the SM, 
 there exist some deviations from the SM expectations,
 namely, in its signal strength of the diphoton decay mode.
Since the Higgs-to-diphoton coupling arises 
 at the quantum level even in the SM, 
 there is a good chance for certain new physics effects 
 to alter the coupling. 
This has motivated many recent studies \cite{AHC-SUSY, AHC-NonSUSY}. 
For studies in this direction before the Higgs boson discovery, 
  see, for example, \cite{AHC-old}.

In our model, the charged scalar $S$ couples 
 to the Higgs boson through the term:
\begin{eqnarray}
{\cal L} &\supset& - \kappa (S^{\dagger}S)(H^{\dagger} H) .
\end{eqnarray}
Hence a one-loop diagram involving $S$ gives 
 a new contribution to the effective Higgs-to-diphoton coupling. 
The amplitude of the loop diagram of $S$ is given by \cite{hgamgam}
\begin{eqnarray}
{\cal A}_S = C \ \kappa \ \frac{v^2}{m_S^2} \ 
 \frac{1}{x_S^2} \left[ -x_S + f(x_S) \right] \ ,
\end{eqnarray}
 while the amplitudes in the SM 
 via the top quark loop and the $W$-boson loop are, respectively,
 given by
\begin{eqnarray}
{\cal A}_t &=& C \ N_c \ Q_f^2  \frac{2}{x_t^2} 
 \left[ x_t + (x_t-1) f(x_t) \right] \ ,
\\
{\cal A}_W &=& -C \ \frac{1}{x_W^2} 
 \left[ 2 x_W^2 + 3x_W + 3 (2x_W-1) f(x_W) \right] \ ,
\end{eqnarray}
 where $C$ is a common constant, $v=174$ GeV, $N_c=3$, $Q_f=2/3$, 
 $x_S \equiv m_h^2/4 m_S^2$, $x_t \equiv m_h^2/4 m_t^2$, 
 and $x_W \equiv m_h^2/4 m_W^2$. 
The function $f(x)$ is defined as 
\begin{eqnarray}
f(x) &\equiv& \arcsin^2 (\sqrt{x})
\end{eqnarray}
 for $x > 1$.

We define the ratio of the branching fraction 
 ${\rm Br}(h \rightarrow \gamma \gamma)$ over its SM value as 
\begin{eqnarray}
r_{\gamma \gamma} \equiv 
\frac{{\rm Br}(h \rightarrow \gamma \gamma)}
{{\rm Br}(h \rightarrow \gamma \gamma) \vert_{\rm SM}}
 = \left\vert 
  1+ \frac{{\cal A}_S}{{\cal A}_t + {\cal A}_W}
  \right\vert^2 \ . 
\label{ratioBr}
\end{eqnarray}
Note that as long as 
 $\vert {\cal A}_S \vert < \vert {\cal A}_t + {\cal A}_W \vert$, 
 ${\cal A}_S$ interferes constructively (destructively) 
 with the SM contributions when $\kappa > 0$ ($\kappa < 0$). 
If $\vert \kappa \vert$ is large and/or the scalar $S$ is light, 
 $\vert {\cal A}_S \vert$ can cause 
 a sizable deviation of the ratio 
 from the SM prediction $r_{\gamma \gamma} = 1 $.

\section{Implications to future experiments}
In our analysis, only three parameters ($y$, $\kappa$ and $m_S$) 
 are involved with the fixed mass difference 
 of $m_\chi = m_S-50$ GeV or $m_\chi = m_S-20$ GeV. 
In order to reproduce the measured relic abundance, 
 we have found the relation between $y$ and $m_S$ 
 in Eq.~(\ref{y-ms}), by which only two parameters 
 ($\kappa$ and $m_S$) are left free. 
As a result, both the spin-independent cross section 
 in Eq.~(\ref{sigmaSI}) 
 and the ratio of the Higgs-to-diphoton branching ratio 
 in Eq.~(\ref{ratioBr}) can be obtained as 
 a function of the two free parameters. 
Therefore, for a fixed $m_S$ value, 
 there is a correlation between $\sigma_{\rm IS}$ 
 and $r_{\gamma \gamma}$ through $\kappa$.

Fig.~4 shows the correlation for various values 
 of $m_S$ with the fixed mass difference of $m_\chi = m_S-50$ GeV. 
As $|\kappa|$ is raised, the resultant spin-independent cross 
 section increases and the deviation of $r_{\gamma \gamma}$ 
 from the SM value $r_{\gamma \gamma}=1$ becomes larger. 
In the Figure, we have imposed the condition $|\kappa| < 4 \pi$ 
 as a naive perturbative bound, 
 which gives the upper bound on the cross section 
 and the deviation of $r_{\gamma \gamma}$ 
 for a fixed $m_S$ value. 
We have found that in order for the spin-independent cross section 
 to be within the reach of the future experimental sensitivity 
 $\sigma_{\rm IS} \gtrsim 10^{-47}$ cm$^{-2}$, 
 the deviation of $r_{\gamma \gamma}$ from the SM prediction 
 should remain sizable. 
Therefore, the precise measurement of Higgs boson decay to 
 diphoton at the LHC and the future direct DM detection experiments 
 are complementary to reveal our scenario.

\begin{figure}[t]
  \begin{center}
   \includegraphics[width=120mm]{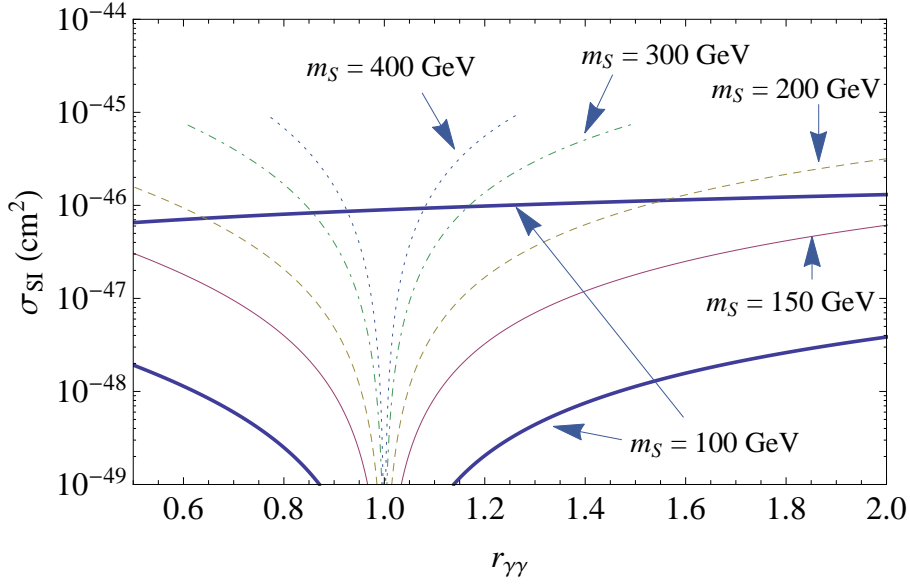}
  \end{center}
 \caption{
 Correlation between $r_{\gamma \gamma}$ and $\sigma_{\rm SI}$ 
 through $|\kappa| < 4 \pi$ with various $m_S$ values.
 The thick solid line, thin solid line, dashed line, 
 dot-dashed line and dotted line, respectively, correspond 
 to the charged scalar mass of $m_S=100$ GeV, 
 150 GeV, 200 GeV, 300 GeV and 400 GeV.
 The DM mass, $m_{\chi}$, is taken as $m_{\chi}=m_S-50$ GeV.
 }
\end{figure}

\begin{figure}[t]
  \begin{center}
   \includegraphics[width=100mm]{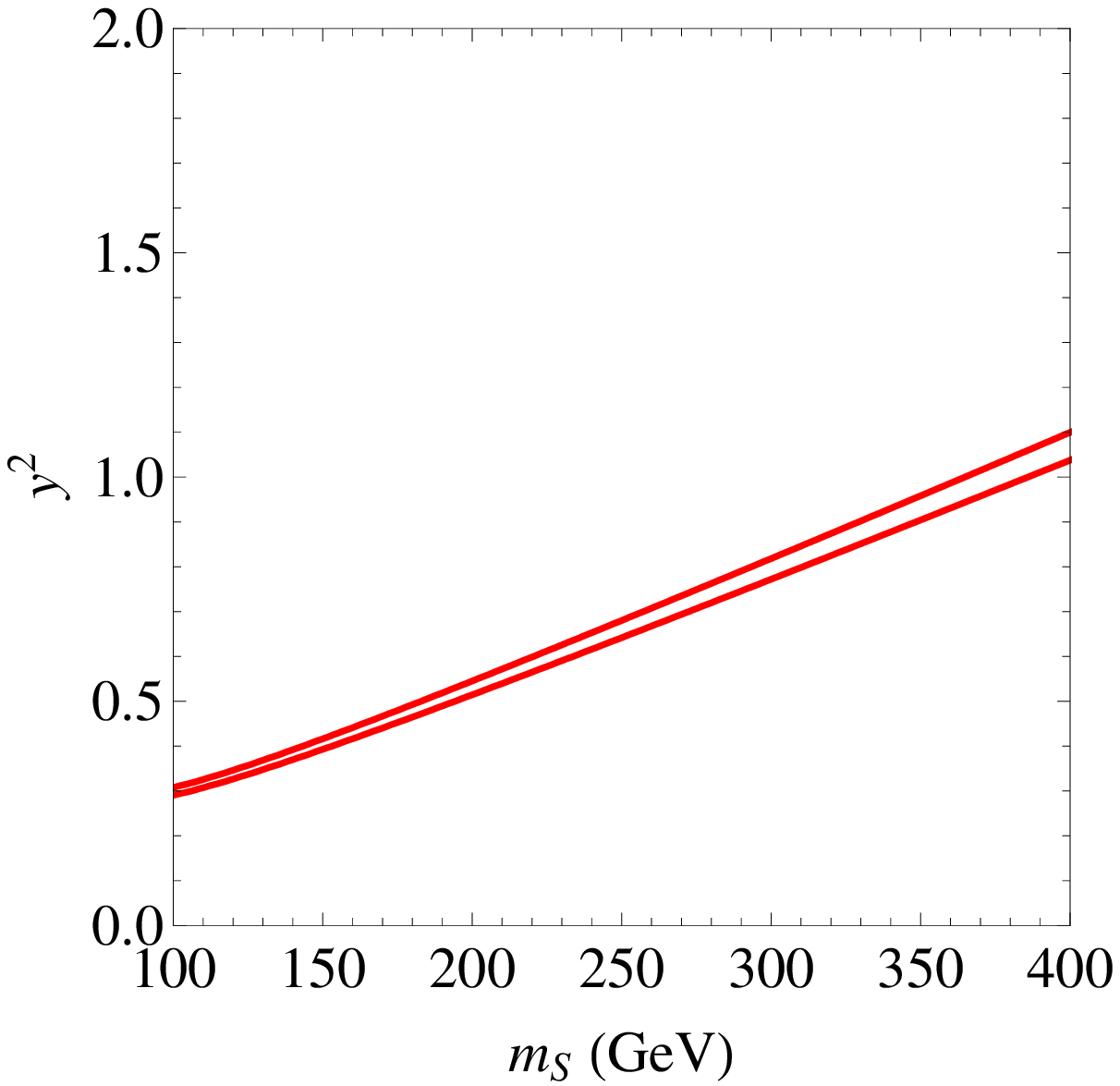}
  \end{center}
 \caption{
Contour plot for the relic abundance of the DM candidate $\chi$
 on the ($m_S$, $y^2$)-plane. 
Here we have taken $m_{\chi}=m_S - 50$ GeV. 
The region between two (red) lines corresponds to 
 the range of the observed DM abundance in Eq.~(\ref{wmap}).
 }
\end{figure}

\section{Case for Dirac fermion dark matter}
In general, we can also consider a model similar 
 to the one described in Sec.~2 except that $\chi$ 
 is a Dirac fermion (see \cite{Agrawal:2011ze} for a similar model). 
In this case, we introduce a global U(1) symmetry, 
 instead of the $Z_2$ parity in Table~1. 
We may assign U(1) charge $+1$ ($-1$) 
 for the Dirac fermion DM (the complex scalar $\chi$). 
The basic Lagrangian remains the same as Eq.~(\ref{basicL}) 
 (except for the correct normalization factor for the Dirac fermion). 
In this section, we will investigate phenomenology 
 for the Dirac fermion case.

Let us first evaluate the relic abundance of $\chi$
 in the same way as in Subsection 2.1 
 but for the Dirac fermion, $\vert {\cal M}(s) \vert^2$ 
 is replaced by
\begin{eqnarray}
\vert {\cal M}(s) \vert^2 &=& 
\frac{1}{g_{\chi}^2} \frac{y^4}{16 \pi} 
 \ 4 \ \frac{1}{2 \sqrt{s - 4 m_{\chi}^2} \sqrt{s}}
\nonumber \\ 
&\times& \left\{ \ \frac{(m_{\chi}^2 - m_S^2)^2}{ m_{\chi}^2-m_S^2 - \frac{1}{2} ( s + \sqrt{s - 4 m_{\chi}^2} \sqrt{s} ) }
- \frac{(m_{\chi}^2 - m_S^2)^2}
 {m_{\chi}^2-m_S^2 - \frac{1}{2} (s - \sqrt{s - 4 m_{\chi}^2} \sqrt{s})} 
\right. \nonumber \\
&+& \left. 2 (m_{\chi}^2 - m_S^2) \ \log 
\left[ \frac{ m_S^2 - m_{\chi}^2 + \frac{1}{2} ( s + \sqrt{s - 4 m_{\chi}^2} \sqrt{s} ) }{ m_S^2 - m_{\chi}^2 + \frac{1}{2} ( s - \sqrt{s - 4 m_{\chi}^2} \sqrt{s} ) } \right]
 + \sqrt{ s - m_{\chi}^2 } \sqrt{s} \ \right\} \ , \nonumber \\
\end{eqnarray}
 with $g_{\chi}=4$. 
Numerically solving the Boltzmann equation, 
 we find the DM relic abundance as a function of $m_S$ and $y^2$. 
Corresponding to Fig.~1, we show our result in Fig.~5, 
 for $m_{\chi}=m_S-50$ GeV. 
As in the case of Majorana fermion DM, 
 the difference between $m_{\chi}$ and $m_S$
 does not significantly alter our results 
 as long as the difference is within $O(10)$ GeV.
 From Fig.~5, we find that the relation 
\begin{eqnarray}
y^2 &\simeq& a (m_S/{\rm GeV} - 100 ) + b \, \label{y-ms2}
\end{eqnarray}
 with 
\begin{eqnarray}
0.0025 \leq a \leq 0.0027 , \ \ \ 0.29 \leq b \leq 0.31 \ ,
\end{eqnarray}
 in order to reproduce the range of the observed 
 relic abundance in Eq.~(\ref{wmap}).

As in the Majorana case, the DM couples to SM quarks 
 at the one-loop level.
In addition to the diagrams in Fig.~2, there is 
 a diagram mediated by photon, shown in Fig.~6, 
 through the magnetic dipole moment induced 
 at one-loop level. 
In fact, this new diagram gives a dominant contribution 
 to the spin-independent cross section 
 for the DM elastic scattering with nucleus\footnote{
For general discussions for a DM particle   
 with electric/magnetic dipole moments, 
 see Ref.~\cite{magnetic dm}. 
}. 
The effective coupling induced by the magnetic dipole moment 
 is found to be 
\begin{eqnarray}
{\cal L}_{\rm eff} &=& 
 (\bar{Q} \gamma^{\mu} Q) (\bar{\chi} \sigma_{\mu \nu} q^{\nu} \chi) 
 \times \left( \frac{-1}{16 \pi^2} \right)
 \ g_{\gamma Q} \ g_{\gamma l} \ y^2 \ \frac{1}{q^2} 
 \ \frac{m_{\chi}}{m_S^2} \ \frac{ x + \log (1-x) }{2 x^2} \ ,
\end{eqnarray}
 where $q^{\mu}$ denotes the momentum transfer 
 from $\chi$ to a quark ($Q$),
 $g_{\gamma l}$ ($g_{\gamma Q}$) denotes 
 the QED coupling of the SM lepton (the SM quark), 
 and $x \equiv m_{\chi}^2/m_S^2$.

The effective charge-charge coupling induced at one-loop level 
 is found to be

\begin{figure}[t]
  \begin{center}
   \includegraphics[width=80mm]{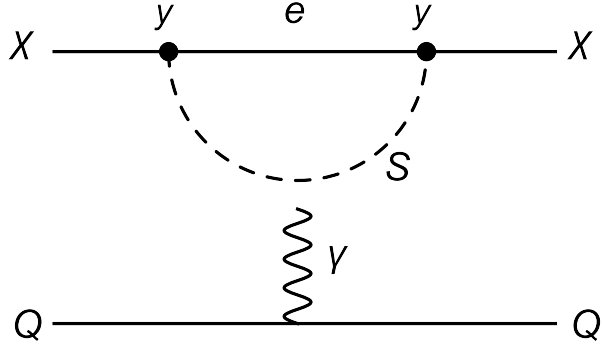}
  \end{center}
 \caption{
 Feynman diagram that gives a dominant contribution 
 to the interaction between $\chi$ and a SM quark 
 when $\chi$ is a Dirac fermion.
 }
\end{figure}

\begin{figure}[t]
  \begin{center}
   \includegraphics[width=120mm]{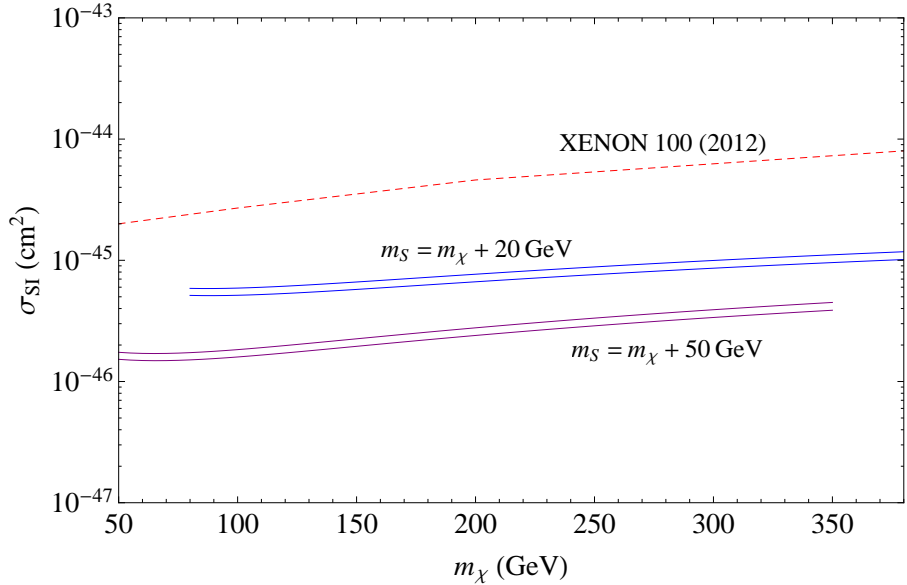}
  \end{center}
 \caption{
 The conservatively-estimated spin-independent cross section for 
  DM-nucleon elastic scattering ($\sigma_{\rm SI}$)
  as a function of the DM mass ($m_{\chi}$).
 The charged scalar mass, $m_S$, is related to the DM mass as
  $m_{\chi}=m_S-50$ GeV (lower two solid lines)
  and $m_{\chi}=m_S-20$ GeV (upper two solid lines).
 The coupling constant $y$ is taken to satisfy 
 the relation in Eq.~(\ref{y-ms2}) 
 so that the model reproduces the observed DM abundance.
 For each pair of degenerate solid lines,
 the upper line corresponds to $(a=0.0027, \ b=0.31)$
 in Eq.~(\ref{y-ms2}), while the lower one to $(a=0.0025, \ b=0.29)$ 
 For comparison, the current experimental bound 
 reported by the XENON 100 experiment \cite{xenon100} 
 is shown by the dashed line.
 }
\end{figure}

Since the vector interactions with each quarks 
 coherently contribute to the total cross section, 
 we obtain the differential cross section 
 for $\chi$-nucleus elastic scattering  
 (in the center-of-mass frame) as 
\begin{eqnarray}
\frac{ {\rm d}\sigma_{\chi-NS} }{ {\rm d}(\cos \theta) } &=&
\frac{1}{8 \pi} \ (Z e)^2 \ g_{\gamma l}^2 
 \left[ \ \frac{2}{1-\cos \theta} + 
 \frac{m_{\chi}^2 -2 m_{NS} \, m_{\chi}}{ (m_{NS}+m_{\chi})^2 }  
 \right] 
\left[\frac{y^2}{16 \pi^2}
 \ \frac{m_{\chi}}{m_S^2} \ \frac{ x + \log (1-x) }{2 x^2} \right]^2 \ ,
\nonumber \\ \label{diff xsec}
\end{eqnarray}
 where $m_N$ and $Z$ respectively denote the mass and 
 the atomic number of a nucleus.
The divergence at $\theta=0$ is cut off by the low energy threshold 
 for a recoiling nucleus. 
To make a conservative comparison of the cross section 
 in our model with the bounds reported 
 by direct DM detection experiments\footnote{
 Direct DM detection experiments actually put a bound on
 the normalization for the distribution of the event 
 rate per recoil energy, 
 d$R/$d$E_r$ ($R$ denotes the event rate and $E_r$ the recoil energy),
 with the assumption that d$R/$d$E_r \propto \exp(-A E_r)$, 
 where $A$ is some known constant.
In our case, however, we have d$R/$d$E_r \propto (1+B/E_r) \exp(-A E_r)$
 with a constant $B$ because of the term $1/(1-\cos \theta)$ 
 in Eq.~(\ref{diff xsec}).
Thus, we cannot naively compare the elastic scattering 
 cross section in our model with an experimental bound.
All we can do is to make a conservative estimate on the cross section by
 substituting $-1$ into $\cos \theta$,
 and compare it with the experimental bound.
},
 we replace $\cos \theta \to -1$ in the right hand side 
 of Eq.~(\ref{diff xsec}). 
Then the conservatively-estimated total cross section 
 for $\chi$-nucleus elastic scattering is given by
\begin{eqnarray}
\bar{\sigma}_{\chi-NS} &=& \int_{-1}^{1} {\rm d}\cos \theta \ 
\frac{ {\rm d}\sigma_{\chi-nucleus} }{ {\rm d}(\cos \theta) } 
\left\vert_{\cos \theta=-1} \right. \nonumber
\\
&=& \frac{1}{4 \pi} \ (Z e)^2 \ g_{\gamma l}^2 \ \left[ \ 1 \ + \ 
\frac{ -2 m_{NS} \, m_{\chi} + m_{\chi}^2 }{ (m_{NS}+m_{\chi})^2 } \ \right] 
\left[ \ \frac{1}{16 \pi^2}
 y^2 \ \frac{m_{\chi}}{m_S^2} \ \frac{ x + \log (1-x) }{2 x^2} \ \right]^2 \ .
\end{eqnarray}
We finally translate the DM-nucleus elastic scattering cross section
 into the spin-independent cross section for DM-nucleon elastic scattering, 
 ($\sigma_{SI}$) as follows:
\begin{eqnarray}
\sigma_{\rm SI} &=& \bar{\sigma}_{\chi-NS} 
 \times 
 \frac{m_N^2}{m_{NS}^2} 
 \frac{(m_{NS}+m_{\chi})^2}{(m_{N}+m_{\chi})^2} \ \frac{1}{A^2} \ ,
\end{eqnarray}
 where $A$ denotes the mass number of the nucleus.
For the XENON 100 experiment, we take $Z=54$ and $A=131.3$.

In Fig.~7, we show the conservatively-estimated
 spin-independent cross section for $\chi$-nucleon elastic scattering,
 $\sigma_{\rm SI}$, as a function of $m_{\chi}$ 
 which is taken to be $m_{\chi}=m_S-50$ GeV and $m_{\chi}=m_S-20$ GeV,
 respectively. 
Here, $y$ is fixed to satisfy the relation of Eq.~(\ref{y-ms2}). 
We see that the resultant cross sections are about an order 
 of magnitude smaller than the current upper bound and will 
 be tested in future direct DM detection experiments.

\section*{Acknowledgments} 

This work is supported in part 
 by the DOE Grant No. DE-FG02-10ER41714 (N.O.),  
 and by the grant of the Japan Society 
 for the Promotion of Science, No. 23-3599 (T.Y.).


\end{document}